\newif\ifcompileiliad 
\compileiliadfalse 

\ifcompileiliad
	\documentclass[aps,prb,preprint,12pt]{revtex4}
	\usepackage[papersize={122mm,160mm},
				hmargin=0.2cm,
				vmargin=0.2cm,
				tmargin=1.0cm,
				bmargin=0.2cm,
				dvips]
				{geometry}
	\usepackage{iwona}
\else
	\documentclass[showpacs, preprint, amsmath, amssymb, superscriptaddress, a4paper]{revtex4}
\fi

\pdfoutput=1

\usepackage{bm}
\usepackage{amssymb}
\usepackage{graphicx}
\usepackage{amsmath}
\usepackage{color}

\newcommand{\bra}[1]{\ensuremath{\left<#1\right|}}
\newcommand{\ket}[1]{\ensuremath{\left|#1\right>}}
\newcommand{\up}{\chi^{\uparrow}}
\newcommand{\down}{\chi^{\downarrow}}
\newcommand{\func}[2]{\phi_{#1}(\bm{r_{#2}})}

\begin{document}

\title{Quantum control theory for coupled 2-electron dynamics in quantum dots}

\author{R. Nepstad}
 \affiliation{Department of Physics and Technology, University
 of Bergen, N-5007 Bergen, Norway}

\author{L. S\ae len}
 \affiliation{Department of Physics and Technology, University
 of Bergen, N-5007 Bergen, Norway}
 \affiliation{Universit\'{e} Pierre et Marie Curie \textendash ~CNRS, Paris, France.}

\author{I. Degani}
 \affiliation{Department of Mathematics, University
 of Bergen, N-5008 Bergen, Norway}

\author{J. P. Hansen}
 \affiliation{Department of Physics and Technology, University
 of Bergen, N-5007 Bergen, Norway}

\begin{abstract}
We investigate optimal control strategies for state to state transitions in a model of a quantum dot molecule containing two active strongly interacting electrons. The Schr{\"o}dinger equation is solved nonperturbatively in conjunction with several quantum control strategies. This results in optimized electric pulses in the THz regime which can populate combinations of states with very short transition times. The speedup compared to intuitively constructed pulses is an order of magnitude. We furthermore make use of optimized pulse control in the simulation of an experimental preparation of the molecular quantum dot system. It is shown that exclusive population of certain excited states leads to a complete suppression of spin dephasing, as was indicated in Nepstad \textit{et al.} [Phys. Rev. B {\bf 77}, 125315 (2008)].
\end{abstract}


\maketitle

\section{Introduction}
Application of quantum control theory to optimize transitions in strongly interacting quantum systems
is a well established technology in simple two-level systems~\cite{two-level}. 
In more complex or open systems involving many important channels it is in general much 
more complicated to improve the probability of desired reactions and transitions. 
This is true in systems as diverse as dipole blockade dynamics in cold Rydberg gases~\cite{rydberg} 
and electron dynamics in semiconductor two-electron quantum dot systems~\cite{saelen}. 
In the latter, which is our case, the electron-electron interaction is comparable to 
other interactions in the system and cannot be neglected. 
The ability to achieve fast and optimized transitions in such systems, and a variety of others, 
is important for improving present day technology in quantum information, metrology and in 
quantum chemistry.

In few-electron quantum dots it is well recognized that interactions with the substrate will induce decoherence, which limits the ability to utilize unique quantum properties such
as entanglement. Examples of such interactions are hyperfine and spin-orbit interactions between the quantum 
dot electrons and surrounding atoms, and interactions with phonons in the substrate lattice. 
As strategies to reduce decoherence, one can either carry out experiments in systems
and at temperatures which minimize unwanted interactions, or try to develop methods to perform
the required transitions much faster than the characteristic timescale of the decoherence.
We have previously demonstrated that intuitively selected microwave pulses can populate both single 
states and more complex states of the lowest excitation bands, and we were able to further 
decrease the transition time in the first case by optimal pulse control.~\cite{saelen}

In the present work we optimize time-dependent transitions to more complex target states and compare various strategies
 of optimization including frequency-selective control algorithms.~\cite{degani}
We show that more advanced control strategies lead to a factor of $7$ faster transition times than
previously reported using intuitively constructed pulses. In the second part we address the application
 of quantum control inside regions of anticrossings. This is related to a recent experiment by 
 Petta \textit{et al.},~\cite{laird} which measures spin dephasing of the system through hyperfine 
 interactions with the surrounding nuclear spin bath. The experiment was simulated in Nepstad \textit{et al.},~\cite{nepstad} 
 and very good agreement between theory and experiment was achieved. In the same work, we further demonstrated how populating higher excited states could be used as a method to inhibit decoherence. In this paper we apply the technique of optimal control theory to exclusively populate such states during initial setup of the experiment. The following section describes the theory in detail. In section III we present the results, followed by concluding remarks.

\section{Theory and method}
In this section we review and detail the numerical methods used to study dynamics of a two-dimensional, two-electron double dot exposed to electric and magnetic fields. This includes DC and pulsed electric fields, strong external magnetic fields and weak, locally varying magnetic fields representing the hyperfine interaction
.~\cite{saelen,nepstad,popsueva}

%
%
\subsection{Model}
The two-dimensional single-particle effective mass Hamiltonian of our system reads
\begin{equation}
 h_0\left( x,y \right) = -\frac{\hbar^2}{2m^*}\nabla^2 + \frac{1}{2} m^*\omega^2 \left[ \left(|x| - \frac{d}{2}\right)^2 + y^2 \right].
  \label{eq:hamiltonian2D}
\end{equation}
Combined with the electron-electron interaction term, the total field-free Hamiltonian becomes
\begin{equation}
 H_0 = h_0({\mathbf{r}_1}) + h_0({\mathbf{r}_2}) + \frac{e^2}{4\pi\epsilon_r\epsilon_0r_{12}}.
 \label{eq:hamiltonian}
\end{equation}
In addition, we include external magnetic and electric fields,
\begin{eqnarray}
  h_{ext}(x,y,t) &=& \frac{e^2}{8m^*}B_{ext}^2(x^2+y^2)+\frac{e}{2m^*}B_{ext}L_z\nonumber\\
 & + & \gamma_eB_{ext}S_z - e\xi(t)x,
  \label{eq:hamiltonian_ext}
\end{eqnarray}
and define the total field Hamiltonian as $H_{ext} = h_{ext}({\mathbf{r}_1}) + h_{ext}({\mathbf{r}_2})$. Here $\mathbf{r}_{1,2}$ are single-particle coordinates, $\xi$ is an electric time dependent field applied along the inter-dot axis and $B_{ext}$ is an external magnetic field perpendicular to the dot. The material parameters may take on different values to reflect various physical systems. In this paper we will use values compatible with GaAs quantum dots, where $m^* = 0.067m_e$ (effective mass), $\epsilon_r = 12.4$ (relative permittivity), $\gamma_e = g^*\frac{e}{2m_e}$ (gyromagnetic ratio) and $g^* = -0.44$ (effective g-factor). The electron mass is denoted $m_e$. The confinement strength is set to $\hbar\omega = 1$~meV and the interdot separation to $d=130$~nm, which are realistic experimental values.~\cite{zumbuhl:256801,petta1}

We obtain eigenstates of the field-free two-electron Hamiltonian $H_0$ by Arnoldi iterations\cite{arpack} using a basis of symmetrized products of one-electron harmonic oscillator functions, $\phi_i$, $\Psi = \sum_{j\ge i}^{n_{max}} c_{ij}|ij\rangle$, where
\begin{equation}
    	|ij\rangle = 
   \begin{cases}
   \frac{1}{\sqrt{2}}[\phi_i({\bf r}_1) \phi_j({\bf r}_2) \pm \phi_j({\bf r}_1) \phi_i({\bf r}_2)] & i \neq j \\
    \phi_i({\bf r}_1) \phi_j({\bf r}_2) & i = j,
    \end{cases}
  \label{twoelbasis}
\end{equation}
with $i,j = \{n_x,n_y\}$ representing the quantum numbers of the Hermite polynomials in $x$ and $y$ respectively. The symmetric and antisymmetric basis functions correspond to singlet and triplet states respectively. We obtain converged results in all cases using $n_{y,max}=4$ and $n_{x,max} = 14$. 

%
%
\subsection{Dynamics in the eigenstate basis}
In cases where the hyperfine interaction between the
two active electrons and the semiconductor nuclei surrounding the quantum dot can be neglected, the
total spin is a conserved quantity. We then need only consider the subspace of symmetric basis functions, corresponding to singlet states, choosing the $+$ sign in Eq.~(\ref{twoelbasis}). The dynamics is governed by the time evolution of the expansion coefficients,
\begin{equation}
	i\dot{c}_{ij}(t) = \sum_{i\prime j\prime} c_{i\prime j\prime}(t)
\langle i\prime j\prime | H | ij \rangle. \label{eq:coeff}
\end{equation}
This system of equations is then integrated using an adaptive form of Adam's method.~\cite{shago} In the singlet subspace using a basis of $\sim 4000$ states, the calculations are reasonably fast. A considerable speedup may be obtained by switching to the eigenstate basis. In this case propagation times of nanosecond duration is performed in less than a minute (on a dual core AMD Turion 64 bits processor). We find converged results using a basis of $50$ eigenstates. The coefficients in Eq.~(\ref{eq:coeff}) become the coefficients of the eigenstates 
\begin{equation}
	i\dot{d}_k(t) = \sum_l d_l(t)
	\langle l | H_{ext} | k \rangle + E_k\,d_k(t),
\end{equation}
where now $|l\rangle$ indicates eigenstate $l$ with corresponding energy $E_l$. The corresponding field matrix elements are calculated using analytic expressions obtained in the harmonic oscillator basis. The required matrix elements are given in detail in Popsueva \textit{et al.}~\cite{popsueva}

%
%
\subsection{Hyperfine interactions}
A particular source of decoherence in double quantum dot molecules is 
the hyperfine interaction with the surrounding substrate nuclei, which has
a characteristic timescale of a few nanoseconds.~\cite{taylor07} To study this interaction from first
principles spin couplings for $\sim 10^6$ nuclear spins surrounding the electrons must be included. 
The magnitude of the interaction is consistent with a random magnetic field of a few~mT. For the timescale of the experiment (~50ns) this is taken to be constant and its spatial dependence can to a good approximation be represented by a step function,~\cite{nepstad}
\begin{equation}
  {\bf B}_N =
  \begin{cases}
    \left( B_x {\bf e}_{x} \ + \ B_y {\bf e}_{y} + \ B_z {\bf e}_{z}\right) & \text{, for } x \ge 0 \\
    0 & \text{, otherwise. }
  \end{cases}
  \label{eq:Bn}
\end{equation}
The hyperfine interaction term is then given by
\begin{equation}
 H_{N} = \gamma_e \sum_{i=1,2} {\bf S}_i \cdot {\bf B}_N \label{HSNC},
\end{equation}
where ${\bf S}_i$ is the spin operator of electron $i$. In this semiclassical picture, we must consider an ensemble of quantum dot systems, each with a different random nuclear magnetic field, and average over the ensemble to obtain physical quantities. To obtain the ensemble, we use a normal distribution of magnetic fields about zero, $ P(\bm{B}_N) = 1/(2\pi B^2_{\rm{nuc}})^{\frac{3}{2}} \exp\left(-\bm{B}_N\cdot\bm{B}_N/2B^2_{\rm{nuc}}\right)$.~\cite{loss} $B_{\rm nuc}$ can be determined by experiments and is of the order of $1$~mT.~\cite{petta1} The interaction term induces couplings between the singlet and triplet states and between the different triplet states, necessitating the inclusion of both subspaces in the calculations. Details of the matrix elements involved can be found in Appendix A. We remark that other types of interactions with external degrees of freedom, such as interactions with electron spins or phonons, can be introduced formally in the same way.

%
%
\subsection{Dynamics in the adiabatic basis}
\label{adiabatic dynamics}
\begin{figure*}[ht]
\begin{center}
	\includegraphics[width=.7\textwidth]{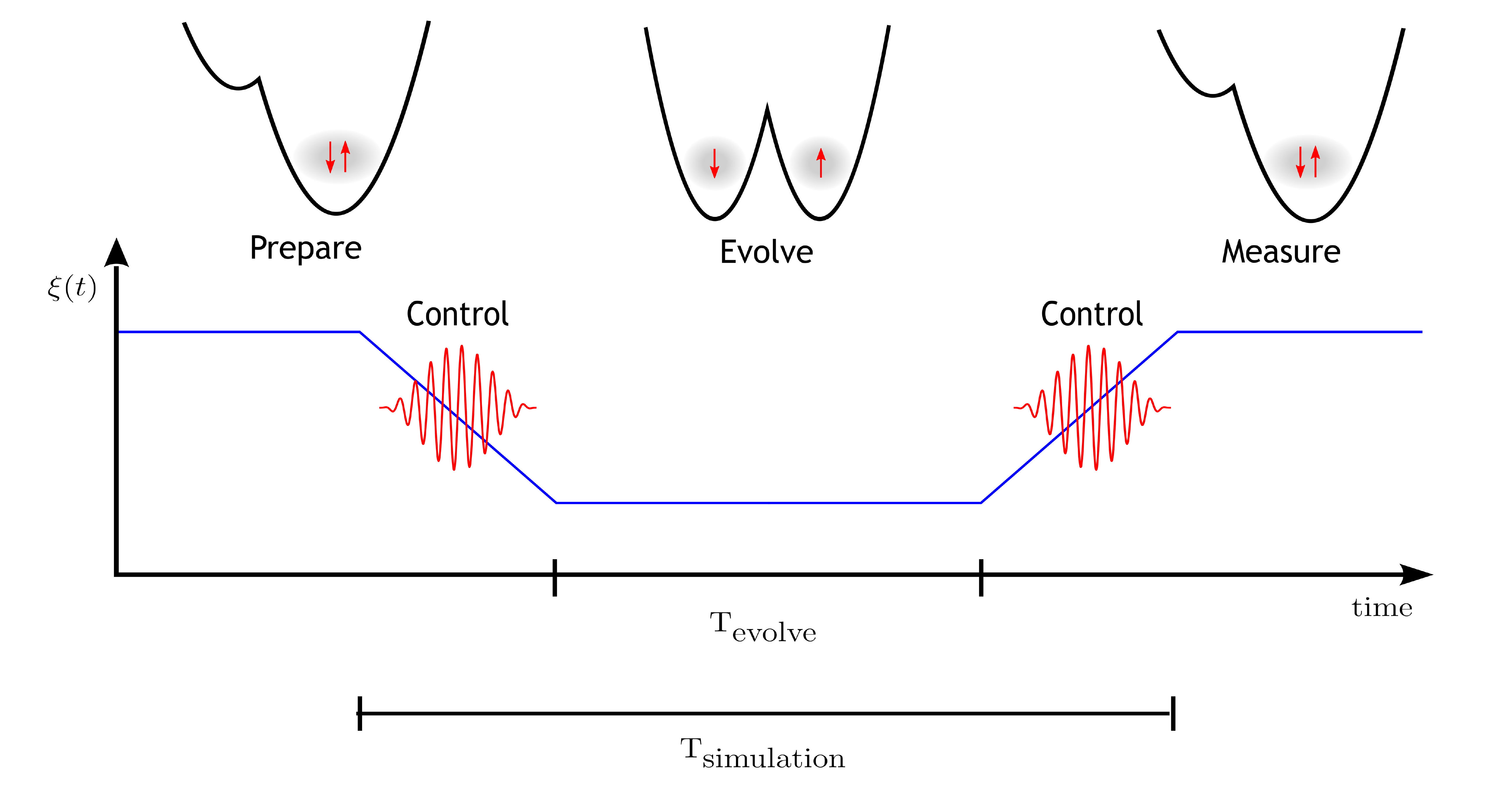}
\end{center}
\caption{{Illustration of a spin dephasing experiment in a double quantum dot molecule, as we model it in our simulations. The blue (dark) line shows the slowly switched electric field that guides the two electrons from a single- to a double-well configuration. During the switch control pulses may be applied in order to guide the electrons into excited states.}}
\label{fig:illustration1}
\end{figure*}

Experimental studies of spin dephasing in quantum dots require preparation of two electrons in the singlet ground state. In a recently reported experiment by Petta \textit{et al.},~\cite{laird} this was achieved by applying a large external electric field to the double dot, deforming the confining potential until at large field strength it became essentially a single dot, as illustrated in Fig.~\ref{fig:illustration1}. The electrons were then allowed to tunnel into the trapping region, forming a singlet state. Reversing the electric field slowly guided the electrons into the ground state of a delocalized double-well configuration, where dephasing occurs. Final readout was performed by once more tuning the electric field to a single dot configuration.

To simulate such an experiment, the one-center basis approach as described above is unsuitable, as a very large number of basis states is required to accurately represent the wavefunction when the electric field is large. Including the triplet states adds an additional factor of four to the basis size, making the calculations prohibitively time consuming. Even switching to the diabatic basis, as described above, yields lengthy calculations and convergence in terms of basis size is arduously obtained. However, we observe that the energy spectrum as a function of electric field strength displays well-separated states with clear anticrossings. These considerations lead us to consider instead an adiabatic basis approach, where the wavefunction is expanded in eigenstates depending parametrically on the electric field,
\begin{equation}
	\Psi (\mathbf{r}_1,\mathbf{r}_2,t) = \sum_{k}c_{k}(t)\theta_k(\mathbf{r}_1,\mathbf{r}_2;\xi) \otimes \ket{S},
	\label{eq:adiabatic_psi}
\end{equation}
where $|S\rangle$ refers to either a symmetric (triplet) or antisymmetric (singlet) spin function. Note that the electric field is time-dependent ($\xi = \xi(t)$), but we have dropped the explicit reference to $t$ in order to simplify notation. The basis states $\theta_k$ are determined from the eigenvalue equation
\begin{equation}
	\left(H_0 - e \xi X\right)\theta(\mathbf{r}_1,\mathbf{r}_2;\xi) = \varepsilon(\xi)\theta(\mathbf{r}_1,\mathbf{r}_2;\xi).
	\label{eq:adiabaticEVP}
\end{equation}
Inserting Eq.~(\ref{eq:adiabatic_psi}) into the TDSE and using Eq.~\ref{eq:adiabaticEVP}, we find the governing equation for the coefficients.
\begin{equation}
\dot c_k(t)\! = -e\dot{\xi} \ \sum_{j \neq k} \frac{\bra{\theta_k} X \ket{\theta_j}} {\varepsilon_k-\epsilon_j}c_j(t) + \imath\varepsilon_k(\xi)c_k(t),
\end{equation}
Written more compactly on vector form, this reads
\begin{equation}
	\dot{\mathbf{c}}(t) = \left(-e\dot{\xi} \mathbf{K}(\xi) + \imath\bm{\varepsilon}(\xi)\right)\mathbf{c}(t).
	\label{eq:ATDSE_vector}
\end{equation}
The antihermitian matrix $\mathbf{K}_{}(\xi)$ is computed for a set of electric field values $\left\{\xi_m\right\}$ using the numerically obtained eigenstates and eigenvalues together with analytic matrix elements of the symmetrized harmonic oscillator functions $\ket{ij}$, defined in Eq.~\ref{twoelbasis},
\begin{equation}
	\mathbf{K}_{kl}^m = \frac{1}{\epsilon_{k}^m-\epsilon_{l}^m}\sum_{ij}\sum_{i\prime j\prime}c_{ijk}^mc_{i\prime j\prime l}^m\bra{ij} X \ket{i\prime j\prime},
\end{equation}
where index $m$ refers to the electric field points.
Since the explicit time dependence in Eq.~(\ref{eq:ATDSE_vector}) is only found in the scalar function $\dot \xi(t)$, the matrix elements need only be computed once, speeding up the time integration. Only $\dot \xi(t)$ must be computed during integration, but this is inexpensive. As the numerically computed basis set is not continuous in $\xi$, we use a simple interpolation between the $\xi$ grid points where required by the integrator.

%
%
\subsection{Optimal control}
\label{subsec:optimalcontrol}
In this section we describe the iterative Krotov method~\cite{krotov} for optimizing optical transitions in quantum systems.~\cite{rasanen} Following we will describe and discuss modifications to this scheme.
In general the method aims to maximize the expectation value of a positive semi-definite operator by means of an external field while minimizing the field energy. The time evolution of the system in which we want to optimize transitions is described by the time-dependent Schr\"{o}dinger equation
\begin{equation}
	i\frac{\partial}{\partial t}\Psi(\bm{r}_1,\bm{r}_2,t) = \left[H_0 - e\epsilon(t)X\right]\Psi(\bm{r}_1,\bm{r}_2,t),
	\label{eq:tdse}
\end{equation}
where $\epsilon(t)$ is an electric field and $X = x_1+x_2$ as before. We have chosen to use $\epsilon(t)$ for the electric field whenever we refer to (optimized) pulses. $H_0$ is the field free Hamiltonian in Eq.~(\ref{eq:hamiltonian}). Our goal is to apply control theory to find optimal pulses for population transfer from an initial state $\Phi_i = \Psi(t=0)$ to a target state $\Phi_t$. The states are preselected and the pulse duration is fixed to $t=T$. The optimization is done by maximizing the expectation value of a projection operator, $|\Phi_t\rangle\langle\Phi_t|$, that is, maximizing the functional $J_1[\Psi] = \langle\Psi(T)|\Phi_t\rangle\langle\Phi_t|\Psi(T)\rangle =|\langle\Phi_t|\Psi(T)\rangle|^2$. The requirement that the field intensity should be as small as possible is achieved by minimizing a second functional, $J_2[\epsilon] = \int_0^Tdt\,\lambda(t)[\epsilon^2(t)]$, where the predefined function $\lambda(t)$ acts as a penalty factor, which can be used to impose an envelope on the electric field. We will use $\lambda(t)$ = $\lambda$ unless otherwise stated. In each iteration, the updated control field is found as a solution to
\begin{equation}
 	\nabla J_a[\epsilon] = 0, \hspace{20pt} J_a = J_1-J_2.
	\label{J_a}
\end{equation}
We proceed to sketch a simple implementation of the Krotov iteration algorithm: The time interval, $[0,T]$, is divided into fixed-length intervals $t_i$ on which $\epsilon(t)$ is taken to be constant, $\epsilon(t_i) = \epsilon_{t_i}$, $t_i \in [0,T]$.
The first step is to integrate the initial value problem of Eq. (\ref{eq:tdse}). For the first iteration, $I=0$, use some initial guess for the control, $\epsilon^0(t_i)$. The choice of initial control is by no means immaterial, as we will see later. After propagating forward, calculate the yield, $|\langle\Phi_t|\Psi^I(T)\rangle|^2$, where $\Psi^I(T)$ is the final state. If the desired yield has been reached, the iterations are terminated. If not, solve the terminal value problem,
\begin{equation}
	\dot{\chi} = -\imath\left[H_0-\mu\epsilon(t)\right]\chi, \hspace{5pt}\mbox{with} \hspace{5pt} \chi(T) = |\Phi_t\rangle\langle\Phi_t|\Psi^I(T)\rangle,
	\label{terminalValueProblem}
\end{equation}
and obtain $\chi^I(t)$. The updated control components $\epsilon^{I+1}_{t_i}$ are obtained while integrating the TDSE, Eq. (\ref{eq:tdse}), again (step-wise) with $\Psi^{I+1}(0)=\Phi_i$: For the first time interval choose $\epsilon^{I+1}_{t_0} = -{\rm{Im}}\langle\chi^I(0)|\mu|\Psi^{I+1}(0)\rangle / \lambda $, and with this $\epsilon^{I+1}_{t_0}$ integrate to find $\Psi^{I+1}(t_1)$. Repeat the process for the next time interval, using
\begin{equation}
	\epsilon^{I+1}_{t_i} = - \frac{ \mbox{\rm{Im}}\langle\chi^I(t_i)|\mu|\Psi^{I+1}(t_i)\rangle}{\lambda}.
	\label{u(t)}
\end{equation}
The entire procedure is repeated until maximum number of iterations or desired yield is reached. Using the eigenstates as basis makes it possible to perform several hundred iterations in a few hours. 

The expression for $\epsilon_{t_i}$, Eq.~(\ref{u(t)}), is a zeroth order approximation to the update equation given to full order in Degani \textit{et al.}~\cite{degani}
Although the simple iteration method described above is not guaranteed to converge monotonically, it works quite well for the system at hand. In fact this feature might even be desirable, as it acts as a 'shake-up' of the numerical calculations: the iteration scheme finds only local maxima and thus adding small perturbations to the solution through having a non-monotonous convergence might lead to even better optimal controls. 
Indeed, this effect was observed when performing an additional update of the control during the backward integration, cf. Eq.~(\ref{terminalValueProblem}). In this case the convergence is smoother but often stagnates around a lower maximum yield. However, we would like to point out that the technique of using backward updates has proven to be quite effective in simpler systems, acquiring extremely high yields.~\cite{zhuRabitz,degani} Tests using a basis of only a few states confirmed this also in our system. 

The method as presented above has another restriction in that it does not discriminate between possible controls, except favoring those of low intensity. This often leads to quite complicated controls that are difficult to produce in an experimental setup. There have been some attempts to address this problem.~\cite{hornung,werschnik} In Werschnik and Gross, a desired structure is enforced by projecting the control onto a preferred subspace in every iteration. Instead of this brute force strategy, a modified functional, $J_b$, can be introduced \cite{degani} which favors low energy controls with a desired structure. This is achieved by choosing a set of `good' controls spanning a desired subspace of the full control space. The `bad' control subspace is then defined to be the orthogonal complement of the `good' subspace.
The weighted terms of the projection of the control onto the `good' and 'bad' subspaces are added to the functional $J_a$. The new functional to maximize is
\begin{eqnarray}
 	 J_b[\epsilon] & = & \langle\Psi(T)|\Phi_t\rangle \langle \Phi_t|\Psi(T)\rangle \nonumber \\
	&- & dt u^T(\lambda I+\lambda_1\Phi_{good}+\lambda_2\Phi_{bad}) u, 
	\label{eq:J_b}
\end{eqnarray}
where the $\lambda$-term is $J_2$, and $\Phi_{good/bad}$ are projection operators onto the `good' and `bad' control subspaces respectively. Here we have used a set of frequencies corresponding to transitions between the $10$ lowest bound states as our space of `good' controls. More specifically the space of `good' controls is defined as the span of $f_{ij} ,\hspace{10pt} i = 0, \hdots , 9, j < i$,
\begin{equation}
      f_{ij}(t_k) = \sin^2 (\pi t_k /T) \cos \left((E_i - E_j )t_k \right),
\end{equation}
$t \in [t_k , t_{k+1} ]$.~\cite{detailCSPM} Optimizing $J_b$ instead of $J_a$ guides the control algorithm in the direction of controls with desired frequencies.

\section{Results}
\begin{figure}[ht]
\begin{center}
	\includegraphics[width=8.6cm]{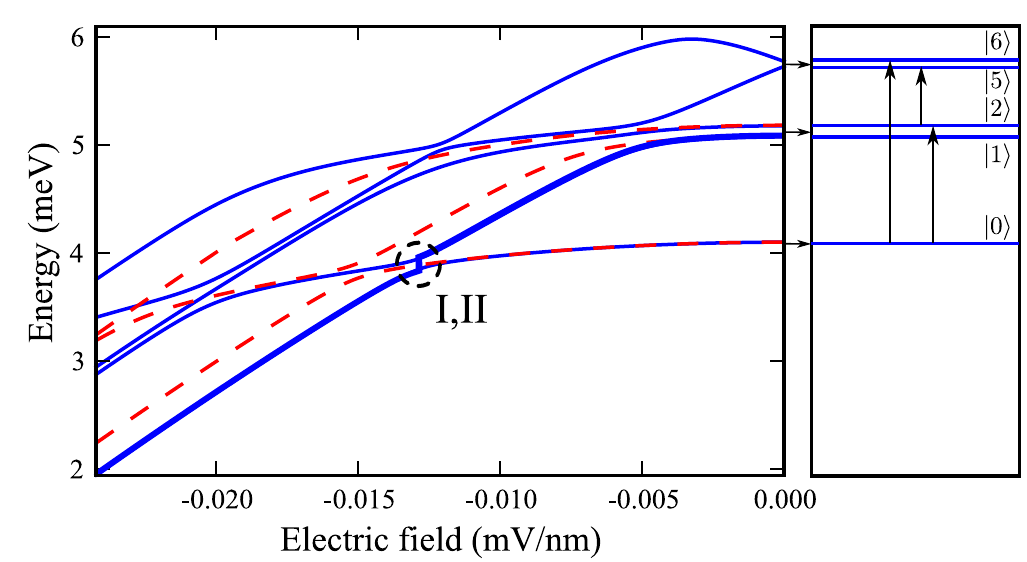}
\end{center}
\caption{{The two-electron double dot spectrum as a function of electric field strength (left), and details of the spectrum at zero electric field (right). Solid lines (blue) are singlet states, dashed lines (red) are triplet states. The arrows indicate transitions referred to in the text. Only states of even $y$-parity are shown.}}
\label{fig:illustration2}
\end{figure}

We here present results of calculations based on the control schemes outlined in the previous section, 
with respect to optimizing simple and combined state to state transitions in the double dot system. 
Section~\ref{Optical dot} deals with transitions in the singlet subspace at zero electric field, 
while section~\ref{Petta dot} focus on transitions during electric gate switching, indicated in Fig.~\ref{fig:illustration2} at the point of anticrossing (dashed black circle). 

The eigenvalue spectrum in Fig.~\ref{fig:illustration2} is shown as a function of electric field strength (left). As an initial strong negative electric field is decreased the state energies are seen to increase linearly and a number of anticrossing regions appear. The physics of the anticrossings normally involves strong state mixing. For example, the ground state in the circled area changes from a state containing essentially both electrons in one dot, to a covalent two-center state as the anticrossing is traversed. The molecular states at zero electric field (right panel) were classified and labeled in Popsueva \textit{et al.}~\cite{popsueva} 

In the present calculations we employ electric fields linearly polarized in the $x$-direction. This couples states which have different $x$-parity and equal $y$-parity. Fig.~\ref{fig:illustration2} shows only states with the same $y$-parity as the ground state. The states $|1\rangle, |2\rangle$ correspond to single exciton states while the states $|5\rangle, |6\rangle$ are ionic states. As the dot separation tends to infinity, these become degenerate and consist of two-electron single dot ground states with both electrons in the same dot.

%
%
\subsection{Spin conserving dynamics in the singlet subspace}
\label{Optical dot}
First, neglecting any spin interactions, we restrict our attention to dynamics in the subspace of singlet states. From the ground state, transitions to the states $\ket{2} \text{and} \ket{6}$ are dipole-allowed, while state $\ket{5}$ can be reached via $\ket{2}$. We will study each of these transitions, finding that optimization procedures can produce very short pulses which achieve almost unit probability transfer.

\subsubsection{Single state transitions}
\begin{figure}[ht]
\begin{center}
	\includegraphics[]{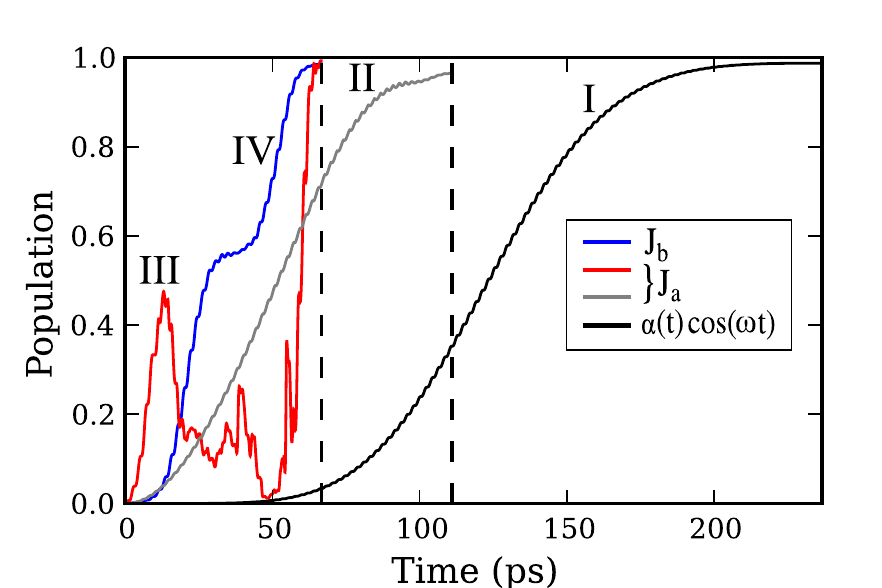}
\end{center}
\caption{Figure showing (optimized) transitions from $\ket{0}$ to $\ket{2}$. Black line (I): ``intuitive pulse'', $\alpha(t)\cos(\omega t)$, where $\omega = (E_2-E_0)/\hbar = 1.5$ THz and $\alpha(t)$ is a $\sin^2$-envelope. Gray line (II): optimized pulse using the functional $J_a$, duration is $111$ ps. Red line (III): optimized pulse using the functional $J_a$, duration is $67$ ps. Blue line (IV): optimized pulse using the functional $J_b$, duration is $67$ ps. The optimization was done using $\Delta t$ = 0.5}
\label{fig:0->2}
\end{figure}

Investigating the $\ket{0}-\ket{2}$ transition, we find that an ``intuitive'' sine-squared envelope pulse tuned to the resonance frequency will transfer $98.7\%$ of the population in $237$ ps. The population of $\ket{2}$ during the pulse is shown in Fig.~\ref{fig:0->2}, labeled I (black curve). In a previous attempt at optimizing this transition, \cite{saelen} we found that $96.5\%$ transfer could be achieved in $111$ ps, using an energy penalty functional and amplitude cutoff (II - gray curve). With the present approach, the same functional ($J_a$) provides better results, transferring $98.6\%$ of the population in only $67$ ps (III - blue curve). Replacing the energy functional $J_a$ with the structure functional $J_b$ gives a slightly better final population of $99.3\%$ (IV - red curve). 
 
With transition time decreased to $67$ ps, the population transfer proceeds in an irregular manner for the energy-penalty optimized pulse, III. During the pulse, as much at $70\%$ of the population is transfered to highly excited states ($>10$). Direct transitions to these states from the ground state may be discouraged by using the structure penalty which will favor the corresponding frequencies and disfavor all others. Switching to the functional $J_b$, the resulting pulse causes population of higher excited states to reduce to $20\%$ (IV). The high numbers are mainly due to the $13$th excited state, which has a strong coupling to the $2$nd excited state. The resonance frequency of this transition is close to that of the $\ket{0}-\ket{2}$ transition, which is included in the space of good controls. Disregarding the population of $\ket{13}$, the population of the remaining higher excited states is $31\%$ and $2.7\%$ for the $J_a$ and $J_b$ optimizations respectively. The plateau structure in the population of $\ket{2}$ during the two short pulses (III and IV) is due to transient population of $\ket{13}$.

\subsubsection{Charge localization}
\begin{figure}[ht]
 \begin{center}
 	\includegraphics[width=8.6cm]{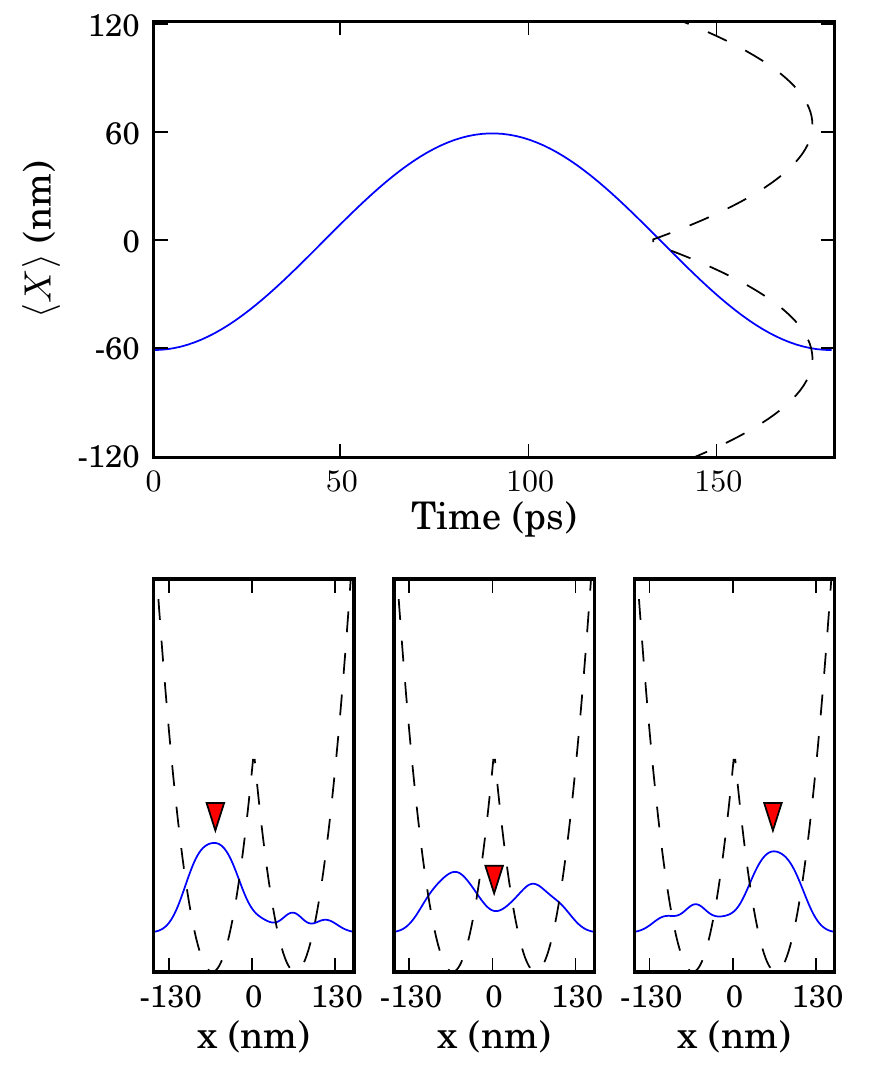}
	\caption{Field-free time evolution of the charge localized state. Upper panel: Time evolution of the expectation value of $X = x_1 + x_2$. Lower panel: Single-electron density averaged over the $y$-coordinate at three different times during the time evolution, $t=0$ ps (left), $t=45$ ps (center) and $t=90$ ps (right). The red markers indicate the value of $\langle X\rangle$ at the three times. The confinement potential is indicated by the dashed lines.}
	\label{chargeLoc}
 \end{center}
\end{figure}

Previously we demonstrated how charge localization in one dot can be achieved in less than a nanosecond by applying weak, resonant pulses on the system.~\cite{saelen} The charge localized state (CLS) is a combination of two states in the third energy band of the spectrum exhibiting ionic structure, in analogy to ionic states in diatomic molecules. At large interdot separation the two states resemble the asymptotic states
\begin{equation}
  |g(r_{1L},r_{2L})\rangle \pm |g(r_{1R},r_{2R})\rangle
\label{leftright}
\end{equation}
where $|g\rangle$ refers to the shifted ground state of a single two-electron dot. Creating an equal linear combination of these states will cause the two electrons to oscillate between localization in the left and the right dot with a period of $180$ ps, inducing a weak current over the dot. This is illustrated in Fig.~\ref{chargeLoc}, where the upper panel shows the expectation value of $X = x_1+x_2$ as a function of time, its value oscillating between the two minima of the double-dot potential. Also shown is the integrated one-electron density of the CLS at certain times during field-free time evolution, $\rho(x) = \int dy_1\ d^2 r_2\ |\Psi (\mathbf{r}_1, \mathbf{r}_2)|^2$.

\begin{figure}[ht]
\begin{center}
	\includegraphics[width=8.6cm]{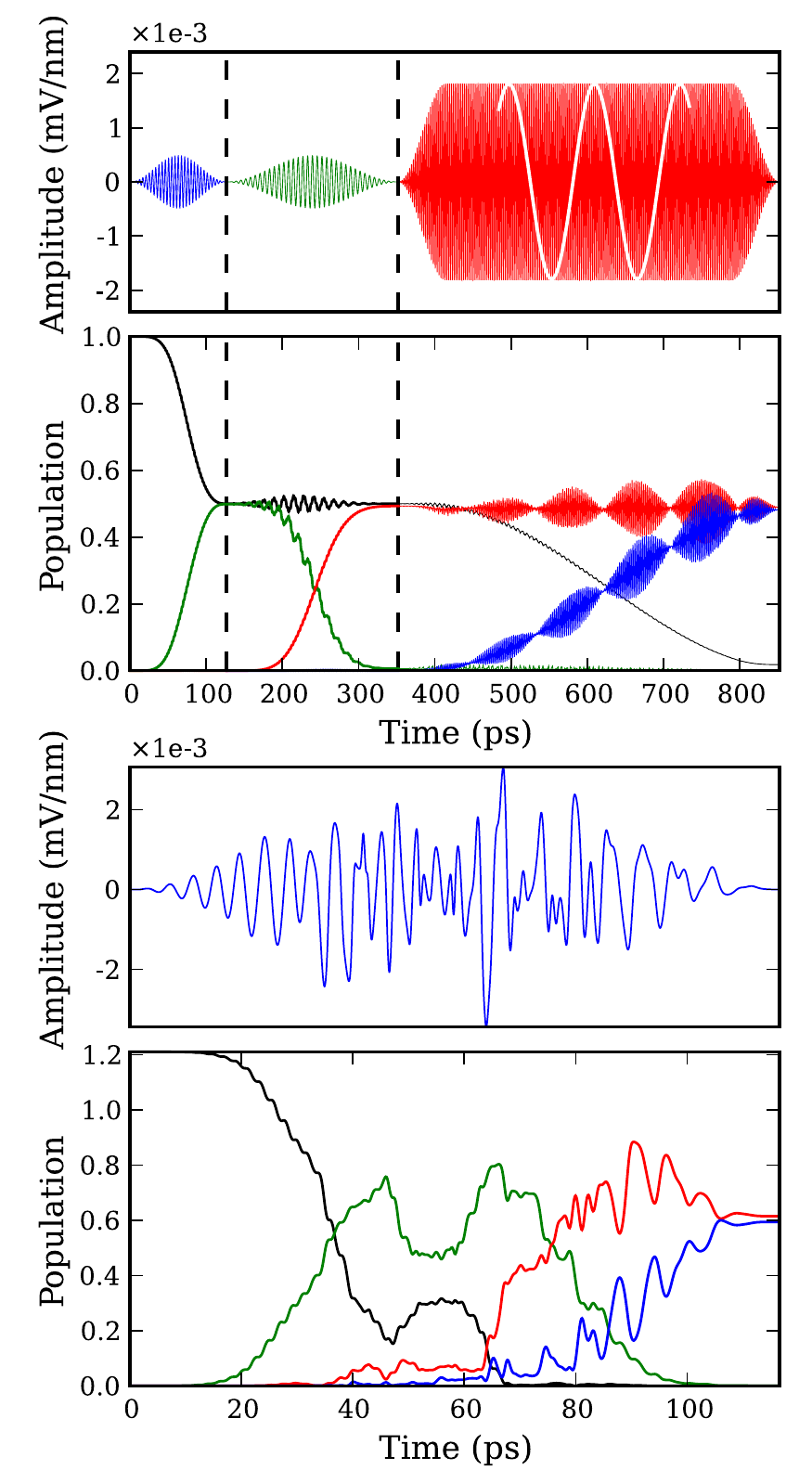}
\end{center}
\caption{Probabilities for the transitions to the charge localized state during the different approaches: (second panel) using a sequence of intuitive pulses~\cite{saelen} and (last panel) optimizing the field using the $J_a$ functional. Populations are plotted for the ground state (black curves), the second excited state (green curves) and the two ionic states (red and blue curves). The corresponding pulses are shown above both panels. The white line in the upper panel is a close-up of the last pulse. The final times are $852$ ps and $117$ ps. The optimization was done using $\Delta t = 0.05$. 
}
\label{fig:comparePRL_CLS}
\end{figure}

In the intuitive scheme the transition to the CLS is achieved via an intermediate transition to the second excited state in the second energy band (labelled $|2\rangle$ in Fig.~\ref{fig:illustration2}). This is necessary because the lowest ionic state has positive $x$-parity and can not be reached from the ground state directly, due to selection rules. The three transitions involved are indicated by arrows in the rightmost part of Fig.~\ref{fig:illustration2}. 
Fig.~\ref{fig:comparePRL_CLS} (second panel from top) shows the eigenstate population as a function of time during the sequence of resonance pulses and during the optimized pulse (bottom panel). The respective pulses are shown above. The first two pulses in the uppermost panel use a $\sin^2$ envelope whereas the last pulse uses a $\sin^2$ ramp-on over $10$ oscillations. 
The optimized pulse was obtained using the functional $J_a$ with final time $T = 117$ ps and a maximum of $300$ iterations. We used $\lambda(t) = 1/\sin^2(\pi\,t/ T)$ to ensure that the pulse is zero at $t=0$ and $t=T$ (note that $\lambda$ is a penalty factor, making the penalty for a non-zero field at the endpoints infinite). As seen from the population during the optimized pulse, the strategy of using the second excited state as an intermediate transition is also being used here, although the transitions are somewhat more involved. The total transition time has been brought down from $852$ ps to $117$ ps using the optimized pulse, and the population of the target state has been improved from $97.2\%$ to $99.8\%$. An important thing to note is that using a defined target state we are also able to selectively choose the configuration of the charge localized state which is determined by the relative phase between the ionic states. We have used a target state defining the two electrons in the left dot at the end of the pulse. The charge oscillations of the CLS has a period of $\sim 180$ ps, and so the relative phase evolution of the ionic states is important during the propagation. 
In this sense the optimal control scheme is stricter compared to the intuitive approach, where we did not control the final configuration of the electrons, only the population in each eigenstate. When considering simple state to state transitions, the phases naturally are not important.
\begin{figure}[ht]
\begin{center}
	\includegraphics[width=8.6cm]{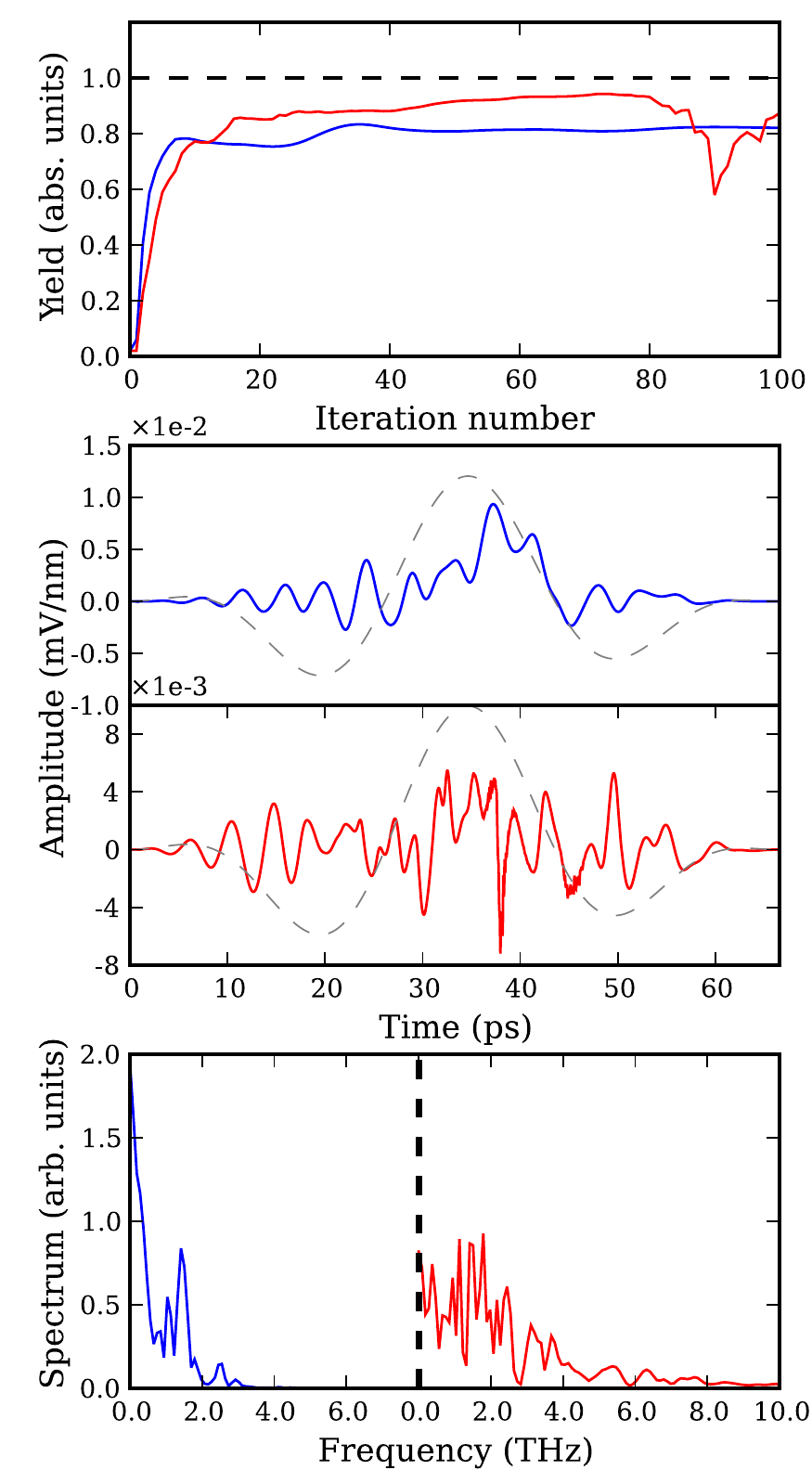}
\end{center}
\caption{Properties of the optimization routines using only energy penalty (red) and including structure penalty (blue) in the transition to the CLS. The upper panel shows the convergence of the yield as a function of iteration number. The middle panel shows the optimal pulses obtained in the two optimizations. The dashed gray curve in the background shows the initial starting field, $0.01\sin^2(\pi\,t/T)\cos(0.01 t)$. The bottom panel shows the spectrum of the two pulses. The final time is $67$ ps and the timestep used is $0.5$.}
\label{fig:Ja_vs_Jb}
\end{figure}
In Fig.~\ref{fig:Ja_vs_Jb} we have applied $J_b$ to optimize transition to the charge localized state. The results for $T=67$ ps are compared with optimization using the $J_a$ functional. Results for $J_b$ and $J_a$ are shown in blue and red curves respectively (all panels). The upper panel shows the convergence of the yield (projection of the final state onto the target state) as a function of iteration. Additional iterations did not produce higher yields. The red curve (upper panel) is fluctuating strongly and the final pulse is also somewhat irregular. In this case we applied a low pass filter to the final pulse to get rid of very high frequency components, caused by numerical noise. We checked that removal of these components did not affect the final yield and dynamics. In general, we experienced greater difficulties in achieving converging results using only the energy penalty, and the yield often converged to zero. The maximum yields for the two methods were $94.3\%$ ($J_a$) and $83.3\%$ ($J_b$). 
While the structure penalty strongly limits the presence of unwanted frequencies in the optimized pulse, population of excited states beyond the 10 lowest still occurs. This is again related to the existence of resonant transitions to higher excited states matching the frequency of the desired transitions. The population of excited states ($ > 13$) is for both methods $\sim 20\%$. 

\begin{figure}[ht]
\begin{center}
	\includegraphics[width=8.6cm]{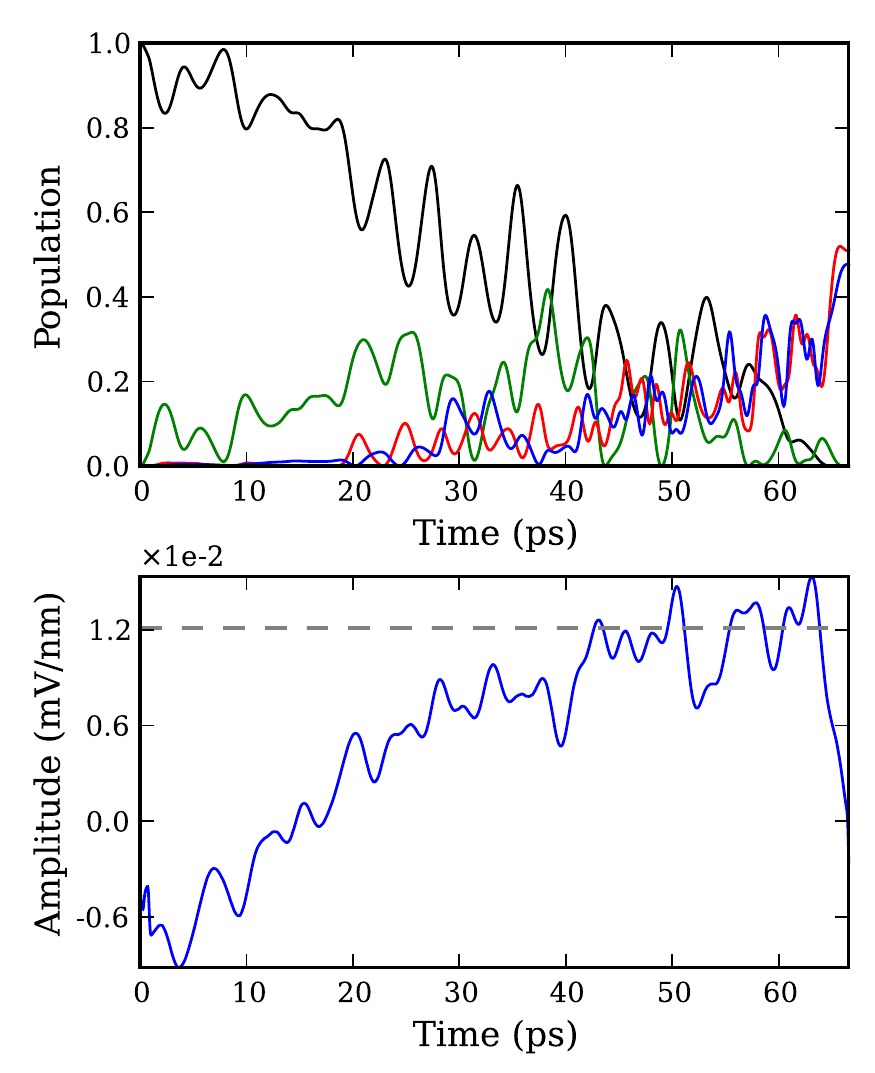}
\end{center}
\caption{{Figure showing an optimized pulse for the transition from $\ket{0}$ to $\ket{CLS}$ using a constant initial field (lower panel). Population as a function of time is plotted in the upper panel for the ground state (black curve), the second excited state (green curve) and the two ionic states (red and blue curve). }}

\label{fig:constInitField}
\end{figure}

We have noticed in all our calculations that the pulse produced by the optimal control algorithm is sensitive to the choice of initial field. An example of this is shown in Fig.~\ref{fig:constInitField}.
Here, we start the iterations using a constant initial field (gray, horizontal curve), and consequently obtain a rather different optimal pulse compared with the one in Fig.~\ref{fig:Ja_vs_Jb} (blue curve), where a $\sin^2$ enveloped pulse was used (gray curve). In this case we see that the optimized pulse has retained much of its initial DC component. The maximum population of the target state is $98.6\%$ after $419$ iterations. In this case the population of highly excited states during the pulse is considerably less, with $<10\%$ in the $24$ highest states. Optimization using only the energy penalty in this case gave a very short, high frequency and high intensity pulse, with a resulting yield of only $\sim 60\%$. 
These examples illustrate the limitations of using only energy penalty when the propagation time becomes short, and how adding structure penalty can consistently guide the control towards a wanted frequency space. 

%
%
We end this section with some comments on the issue of using $y$-polarized fields. The transition to the lowest ionic state (with positive $x$-parity) could in principle have been achieved using a $y$-polarized field and the third excited state in the energy spectrum, which has positive $x$-parity and negative $y$-parity. Note that the transition to the upper ionic state (with negative $x$-parity) can only be reached via $x$-polarized fields. Using this scheme one could perform the two operations simultaneously with weak fields. 

There are however properties of the spectrum obstructing the use of $y$-polarization in this system. In the case of the CLS, the coupling between the third excited state and the lowest ionic state is virtually zero. Moreover, from the lower energy bands, there exist a multitude of strong couplings to states further up in the spectrum, precluding selective transitions to lower lying states. Numerical calculations confirm that selective state population is impractical using $y$-polarized fields.

This feature of the double dot spectrum is related to the symmetry of the potential, particularly in the $y$-direction. As noted in an earlier work,~\cite{saelen} and as we will also see manifested later on, optical manipulation in this two-electron molecule system is actually restricted by the degree of symmetry in the potential, and control would be more easily achieved in slightly asymmetric dots. Similarly, we expect that $y$-polarized fields could be more useful in anharmonic systems.

%
%
\subsection{Optimized transitions and spin interactions}
\label{Petta dot}
In a previous paper,~\cite{nepstad} we studied the effects of spin-dephasing in the quantum dot system, modelling the experiment described in Sec.~\ref{adiabatic dynamics}, and replicating experimental conditions as accurately as possible. We observed that when using ultrafast electric switching ($1$ ps) through the anticrossing (black-dashed circle in Fig.~\ref{fig:illustration2}), large population transfer from the ground state to the second energy band resulted. The decohence is largely suppressed for those states, and when the system was switched back to the 'single dot' configuration, $95\%$ of the initial singlet population was regained. The suppression is explained by the fact that at zero electric field, the singlet-triplet energy splitting is approximately $100$ times greater for the second excited singlet state compared with the ground state. Some of the population vanished to higher excited states during passage through the anticrossings, causing the $5\%$ loss.

\begin{figure}[ht]
\begin{center}
	\includegraphics[width=8.6cm]{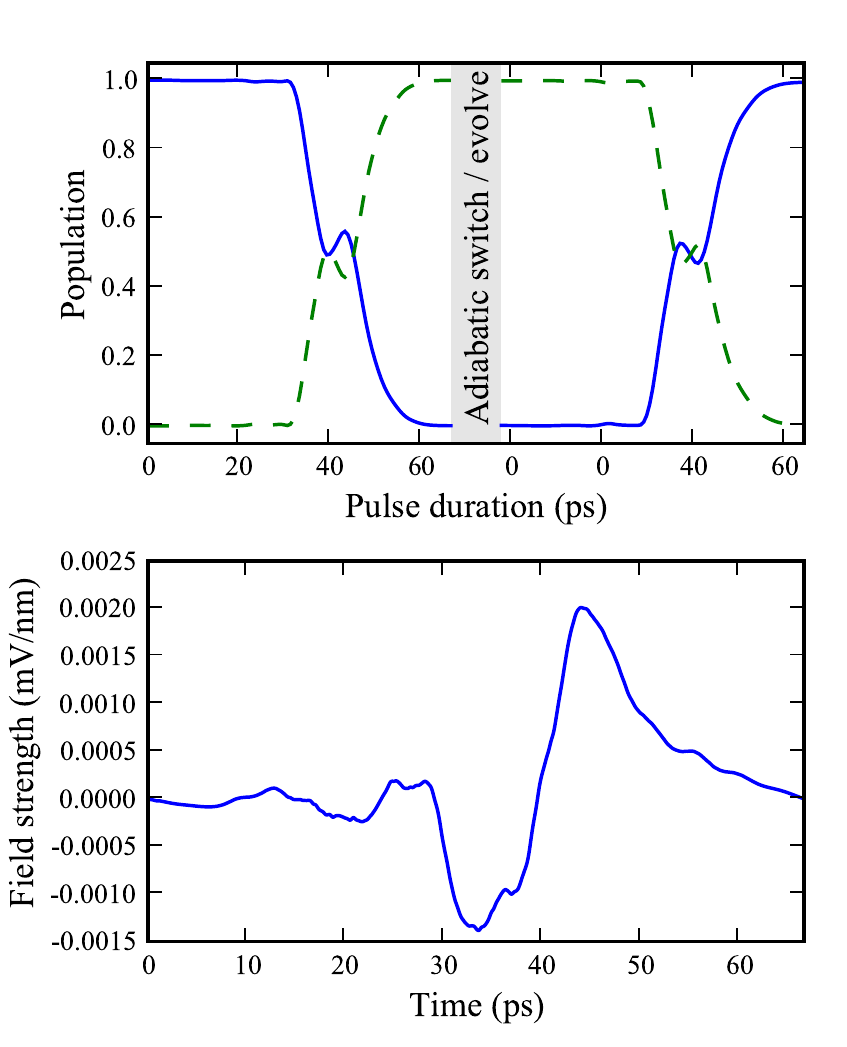}
\end{center}
\caption{Figure showing optimized transition and corresponding pulse at the anticrossing in Fig.~\ref{fig:illustration1}. The upper panel shows population in the two lowest energy eigenstates during the pulse, before and after the adiabatic switch and evolve, indicated by the gray area (see text for description). The same pulse is employed in both cases. Lower panel: the optimized pulse.}
\label{fig:anticrossingTransition}
\end{figure}

By applying optimal control schemes in combination with the adiabatic electric switch, the transition to excited states may be achieved with near $100\%$ probability. An optimized pulse applied at the point of anticrossing will force a non-adiabatic transition and by targeting the desired excited state explicitly we minimize loss to other states. Fig.~\ref{fig:anticrossingTransition} shows such a transition between the two lowest eigenstates using an optimized pulse.
The optimized pulse was obtained using the structure functional $J_b$ together with $\lambda(t) = 1/\sin(\pi t/T)$, and has a duration of $T=67$ ps. In this case the population of other states during the pulse is completely negligible and the final population of the second excited state is as high as $99.9 \%$. After the pulse, an adiabatic switch of duration $2$ ns is applied, guiding the system to the delocalized double-well configuration, where the system is left to interact with the spin bath for $50$ ns. Reversing the adiabatic switch and optimized pulse procedure, we find that $99.3 \%$ of the ground state population is regained.

\section{Summary}
In this work we have demonstrated to what extent quantum control strategies can be 
applied to obtain required transitions between electronic states of two-electron quantum dot
molecules. Such transitions are non-trivial partly due to the strong electron-electron interaction, 
but also the large number of coupled states induced by the external fields. 
Nevertheless, the calculations have shown that single states and superposition of states may be 
reached with close to 100\% probability.

Using weak, pulsed electric fields in the THz regime, we have shown that transitions from the ground state to a preselected excited state may be obtained within ~100 ps. When applying advanced control strategies, a speedup of more than $7$ times the transition time using straightforward intuitive pulses is gained. Such control strategies also have the advantage of returning a final pulse consisting of experimentally relevant frequencies. In the case of interactions with slowly varying external fields, which have been applied in experiments, we have shown that complete transitions at anticrossings can be obtained. This is a realistic implementation of a fully diabatic time development, which in the Landau-Zener model requires infinitely fast transitions. We also showed that the hyperfine interaction in the excited states is unimportant at the considered time scales as opposed to in the ground state. Advanced engineering of tailored pulses as here described appear as a realistic route to accessing and manipulating electronic states in experiments.

\subsection*{Appendix A}
\subsubsection*{Matrix elements for the hyperfine interaction}
\label{App. A}
In this Appendix we give details of the matrix elements for the hyperfine interaction in Eq.~(\ref{HSNC}) using the symmetrized basis of Hermite functions. 
The spin-states are as usual
\begin{equation}
	 \mbox{Triplet} \left\{\begin{array}{ll}
	\up(1)\up(2) & = \ket{T_+}\\
	\down(1)\down(2) & = \ket{T_-}\\
	\frac{1}{\sqrt{2}}\left(\up(1)\down(2)+\down(1)\up(2)\right) & = \ket{T_0}
	    \end{array} \right.
\end{equation}
\begin{equation}
 	\mbox{Singlet} \left\{ \frac{1}{\sqrt{2}}\left(\up(1)\down(2)-\down(1)\up(2)\right)\right. = \ket{S}
\end{equation}
The singlet has corresponding symmetric spatial function of the form
\begin{equation}
	\ket{\Psi_I(\bm{r}_1,\bm{r_2})} \Rightarrow
		\left\{\begin{array}{l}
			 \ket{ii} = \func{i}{1}\func{i}{2} \\
			 \ket{ij} = \frac{1}{\sqrt{2}}\left(\func{i}{1}\func{j}{2}+\func{j}{1}\func{i}{2}\right)
							\end{array} \right.
\end{equation}
The triplet has corresponding antisymmetric spatial function of the form
\begin{equation}
	\ket{\Psi_J(\bm{r}_1,\bm{r_2})} \Rightarrow
			 \ket{kl} = \frac{1}{\sqrt{2}}\left(\func{k}{1}\func{l}{2}-\func{l}{1}\func{k}{2}\right)
\end{equation}
The $\func{i}{}$s are as before two dimensional Hermite functions with $i=\{n_x,n_y\}$. 
Recall also the representation of the effective nuclear field, Eq.~(\ref{eq:Bn}),
\begin{equation}
    {\bf B}_N =
    \begin{cases}
       \left( B_x {\bf e}_{x} \ + \ B_y {\bf e}_{y} + \ B_z {\bf e}_{z}\right) & \text{, for } x \ge 0 \\
        0 & \text{, otherwise. }
    \end{cases}
\end{equation}
\subsubsection*{Matrix elements for the $ | S \rangle \longleftrightarrow | T \rangle$ coupling}
\begin{equation}
	\bra{\Psi_I(\bm{r}_1,\bm{r_2});S}\hat{H}_{N}\ket{\Psi_J(\bm{r}_1,\bm{r_2});T}
\end{equation}
The first case reads
\begin{eqnarray}
	 & & \bra{ii;S}\sum_{i=1,2} \bm{S}_i\cdot\bm{B}_N\ket{kl;T} = \nonumber \\
	 & & \frac{1}{\sqrt{2}}\sum_{x_i=x,y,z}\left\{ \delta_{il}\bra{i}B_{x_i}\ket{k}-\delta_{ik}\bra{i}B_{x_i}\ket{l} \right\} \label{spatial1} \\
     & & \times \bra{S}S_{1x_i}-S_{2x_i}\ket{T},
\end{eqnarray}
and the second possibility is
\begin{eqnarray}
	& &  \bra{ij;S}\sum_{i=1,2} \bm{S}_i\cdot\bm{B}_N\ket{kl;T} = \nonumber \\
 	& & \frac{1}{2}\sum_{x_i=x,y,z}\left\{\delta_{jl}\bra{i}B_{x_i}\ket{k} + \delta_{il}\bra{j}B_{x_i}\ket{k} - \delta_{ik}\bra{j}B_{x_i}\ket{l}-\delta_{jk}\bra{i}B_{x_i}\ket{l} \right\} \label{spatial2} \\
	& & \times \bra{S}S_{1x_i}-S_{2x_i}\ket{T}.
\end{eqnarray}

\subsubsection*{Matrix elements for the $ | T \rangle \longleftrightarrow | T \rangle$ coupling}
The only possibility is,
\begin{equation}
	\bra{\Psi_I(\bm{r}_1,\bm{r_2});T}\hat{H}_{N}\ket{\Psi_J(\bm{r}_1,\bm{r_2});T}
\end{equation}
\begin{eqnarray}
	& &  \bra{ij;T}\sum_{i=1,2} \bm{S}_i\cdot\bm{B}_N\ket{kl;T} = \nonumber \\
 	& & \frac{1}{2}\sum_{x_i=x,y,z}\left\{\delta_{jl}\bra{i}B_{x_i}\ket{k} - \delta_{il}\bra{j}B_{x_i}\ket{k} + \delta_{ik}\bra{j}B_{x_i}\ket{l}-\delta_{jk}\bra{i}B_{x_i}\ket{l} \right\} \label{spatial3}\\
	& & \times \bra{T}S_{1x_i}+S_{2x_i}\ket{T}.
\end{eqnarray}
The spin-coupling elements $\bra{S}S_{1x_i}-S_{2x_i}\ket{T}$ and $\bra{T}S_{1x_i}+S_{2x_i}\ket{T}$ for $x_i \in \{x,y,z\}$ are calculated straightforward using the properties of the spin operators. Their values are listed below for the different cases numbered from $a$ to $j$. 
The spatial matrix elements, Eq.~(\ref{spatial1})-(\ref{spatial2}) and ~(\ref{spatial3}), are composed of simple, separable integrals over the Hermite basisfunctions. Since $B=0$ over the left dot the integration in the $x$-direction runs over half the interval,
\begin{equation}
	\int_0^{\infty}{\rm d}x\,\int_{-\infty}^{\infty}{\rm d}y \,\,H_{n_x}(x)H_{n_y}(y)H_{m_x}(x)H_{m_y}(y)\,e^{-(x^2+y^2)}.
\end{equation}
As the Hermite functions have well defined parity we can use the values of tabulated integrals over the whole interval. 
Denoting the matrices made up by the spatial integrals above, by $\mathbf{S}$, $\mathbf{S-T}$ and $\mathbf{T}$ respectively, we can set up the following matrix,

\begin{table*}[htbp]
	\begin{minipage}{0.5\columnwidth}
		\includegraphics[width=4.5cm]{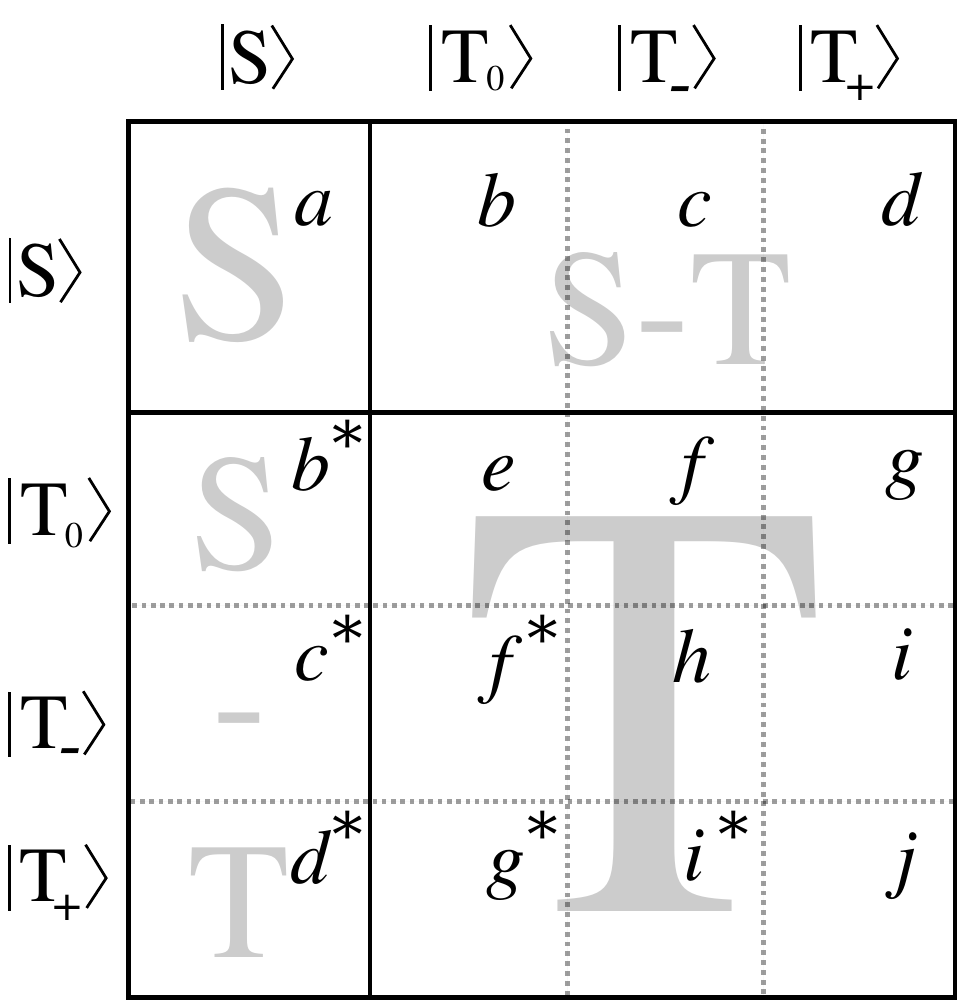}
	\end{minipage}
	\begin{minipage}{0.24\textwidth}
		\begin{eqnarray*}
			a &=& 0\\
			b &=& B_z\\
			c &=& \frac{1}{\sqrt{2}}\left(B_x - iB_y\right)\\
			d &=& -\frac{1}{\sqrt{2}}\left(B_x + iB_y\right)\\
			e &=&0
		\end{eqnarray*}
	\end{minipage}
	\begin{minipage}{0.24\textwidth}
	\begin{eqnarray*}
		f &=& \frac{1}{\sqrt{2}}\left(B_x + iB_y\right)\\
		g &=& \frac{1}{\sqrt{2}}\left(B_x - iB_y\right)\\
		h &=& -B_z\\
		i &=& 0\\
		j &=& B_z
	\end{eqnarray*}
	\end{minipage}
\end{table*}

In the Hermite basis each square in the matrix represents a $\sim 4000\times4000$ matrix. Again we convert to the adiabatic eigenfunction basis at each electric field strength in order to keep the total size of the matrix small ($4n\times 4n, n = 50$).

\end{document}